\documentclass[draft]{agujournal2019}
\usepackage{url} 
\usepackage{comment}
\usepackage[draft=false]{hyperref}
\usepackage{subcaption}
\usepackage{soul}
\usepackage[authoryear]{natbib}
\usepackage{tablefootnote}
\usepackage{microtype}
\usepackage{ragged2e}
\justifying

\draftfalse

\journalname{Space Weather}

\begin{document}

%
%


\title{Correcting Projection Effects in CMEs using
GCS-based Large Statistics of Multi-viewpoint
Observations}

%
%




\authors{Harshita Gandhi\affil{1}, Ritesh Patel$^{2}$, Vaibhav Pant$^{3}$, Satabdwa Majumdar$^{4}$, Sanchita Pal$^{7,8}$, Dipankar Banerjee$^{3,5,6}$, Huw Morgan$^{1}$}


\affiliation{1}{Department of Physics, Aberystwyth University, Aberystwyth, Wales, UK}
\affiliation{2}{Southwest Research Institute, 1050 Walnut Street Suite 300, Boulder CO, USA}
\affiliation{3}{Aryabhatta Research Institute of Observational Sciences, Nainital, India}
\affiliation{4}{Austrian Space Weather Office, GeoSphere Austria, Graz, Austria}
\affiliation{5}{Indian Institute of Astrophysics, 2nd Block Koramangala, Bangalore, India}
\affiliation{6}{Center of Excellence in Space Sciences, IISER, Kolkata, India}
\affiliation{7}{Heliophysics Science Division, NASA Goddard Space Flight Center, Greenbelt, MD 20771, USA}
\affiliation{8}{Department of Physics and Astronomy, George Mason University, Fairfax, VA 22030, USA}




\correspondingauthor{Harshita Gandhi}{hag43@aber.ac.uk}



\begin{keypoints}
\item Presents a large multi-viewpoint dataset containing 3D (true) speeds, 3D (true) widths, locations, and source regions of 360 CMEs observed during the solar cycle 24
\item Statistical analysis of true speed and width of CMEs with and without source region separation is compared to the projected values
\item Corrected speed and width to be used as initial conditions in space weather forecasting models for better arrival time predictions at Earth
\end{keypoints}

%
%

%
%


\begin{abstract}                                                                   
This study addresses the limitations of single-viewpoint observations of Coronal Mass Ejections (CMEs) by presenting results from a 3D catalog of 360 CMEs during solar cycle 24, fitted using the GCS model. The dataset combines 326 previously analyzed CMEs and 34 newly examined events, categorized by their source regions into active region (AR) eruptions, active prominence (AP) eruptions, and prominence eruptions (PE). Estimates of errors are made using a bootstrapping approach. The findings highlight that the average 3D speed of CMEs is $\sim$1.3 times greater than the 2D speed. PE CMEs tend to be slow, with an average speed of 432 km $s^{-1}$. AR and AP speeds are higher, at 723 km $s^{-1}$ and 813 km $s^{-1}$, respectively, with the latter having fewer slow CMEs. The distinctive behavior of AP CMEs is attributed to factors like overlying magnetic field distribution or geometric complexities leading to less accurate GCS fits. A linear fit of projected speed to width gives a gradient of $\sim$2 km $s^{-1}\deg^{-1}$, which increases to 5 km $s^{-1}\deg^{-1}$ when the GCS-fitted `true' parameters are used. Notably, AR CMEs exhibit a high gradient of 7 km $s^{-1}\deg^{-1}$, while AP CMEs show a gradient of 4 km $s^{-1}\deg^{-1}$. PE CMEs, however, lack a significant speed-width relationship. We show that fitting multi-viewpoint CME images to a geometrical model such as GCS is important to study the statistical properties of CMEs, and can lead to a deeper insight into CME behavior that is essential for improving future space weather forecasting.
\end{abstract}

\section*{Plain Language Summary}Space weather refers to the changing conditions in space, largely influenced by massive eruptions from the Sun. We call these eruptions 'Coronal mass ejections' or 'CMEs'. Earth-directed CMEs produce geomagnetic storms that affect our satellites, communication systems, and power grids, causing disruptions in our technology and infrastructure. Hence, how fast CMEs move and how wide they are in 3D space is crucial to predicting their arrival on Earth. To trace these eruptions in 3D, we use a geometrical model on a large set of CMEs during low and high solar activity from different angles, like figuring out the path of a flying bird from different angles.  Derived 3D characteristics such as speed and width (size) compared to the values obtained from one angle (like watching the bird from only one spot) gives us a better idea of how fast they are going. Some eruptions were slow, while others were faster. The bigger the eruption, the faster it tends to be. Our results highlight that the 3D aspect of CMEs is crucial for issuing timely warnings and taking necessary precautions to safeguard our technology and prevent potential damages caused by space weather events.


%
%

%


%
%
%
%

\section{Introduction}

\subsection{Background and Previous Works}

Coronal mass ejections (CMEs) \citep{hundhausen1997coronal} are large-scale structures of plasma and magnetic fields erupting from the Sun at speeds of up to several thousand kilometers per second \citep{webb2012coronal}. CMEs are significant drivers of space weather events on Earth, including geomagnetic storms, auroras, and solar energetic particles (SEPs) \citep{luhman1997cmes,baker1998space,gopalswamy2009coronal}. Auroras are a notable outcome of space weather phenomena, which manifest as luminous displays in the Earth's atmosphere. However, it is important to note that solar energetic particles (SEPs) and geomagnetic storms can lead to substantial disturbances in both terrestrial and space-based equipments \citep{st2000properties,webb2000relationship,wang2002statistical,zhang2003identification,kim2005forecast,michalek2007prediction}, therefore, a crucial component of forecasting CME events at Earth is to understand their kinematics \citep{taktakishvili2009validation,thernisien2006modeling,hutton2017automated} from both a solar \citep{temmer2021space} and a terrestrial perspective \citep{pulkkinen2007space}.

The first CME event was recorded on 14 December 1971 by a coronagraph aboard NASA's Seventh Orbiting Solar Observatory (OSO-7) \citep{tousey1973space}, and led to the realization of the importance of studying CME properties, including their kinematics, based on coronagraph imagery \citep{macqueen1974outer,sheeley1980initial,macqueen1980high}. The Solar and Heliospheric Observatory (SOHO) \citep{domingo1995soho} was launched in 1995, and had a suite of three coronagraphs, called the Large Angle and Spectrometric Coronagraph Experiment (LASCO C1, C2, and C3) \citep{brueckner1995large}, providing a field of view (FOV) from 1.1 to 30 $ R_{\odot}$. LASCO has provided us with more than two decades of data and has significantly bolstered our scientific understanding of CMEs. Since 2006, we have had a stereoscopic picture of CMEs thanks to the advent of the Solar Terrestrial Relations Observatory (STEREO) \citep{kaiser2008stereo} and its two coronagraphs (COR1: FOV of 1.1-4 $ R_{\odot}$ and COR2: FOV of 2-15 $ R_{\odot}$)\citep{howard2008sun}.

Coronagraphs give single-vantage point 2-dimensional images of the Thomson-scattered photospheric light from coronal electrons, integrated along a line of sight through the optically thin corona onto an image plane \citep{burkepile2004role}. Within coronagraph images, CMEs appear as structures that travel outward through the coronagraph FOV \citep{gosling1974mass, hundhausen1993sizes, manchester2017physical}. They have large variations in size, brightness and shapes. Among different CME morphologies, one such morphology is a three-part structure \citep{howard1985coronal} comprised of a bright leading loop, a dark low-density cavity, and a high-density core \citep{hundhausen1993sizes}. This appearance is interpreted as a large magnetic flux rope structure in the corona \citep{vourlidas2014flux, hutton2015}.

Various automatic and manual techniques have been created to detect and track CMEs in \textcolor{black}{a series of coronagraph images} \citep{gosling1974mass,harrison1994statistical,balmaceda2018reliable,munro1979association,macqueen1983kinematics,howard1985coronal,sheeley1986solwind,hundhausen1993sizes, burkepile1993revised, hundhausen1994speeds,st1999comparison,st2000properties,yashiro2004catalog,byrne2012automatic}, and their fundamental characteristics, such as location, speed, angular width, and mass, have been recorded and stored in the catalogs. The most widely used SOHO/LASCO and STEREO/SECCHI CME catalogs are the Coordinated Data Analysis Workshop (\href{https://cdaw.gsfc.nasa.gov/}{CDAW}) LASCO CME catalog (1996-present) \citep{yashiro2004catalog}, the Computer Aided CME Tracking (\href{https://www.sidc.be/cactus/}{CACTus}) LASCO (1997-present) \citep{robbrecht2004automated}, and STEREO (2007 - 2014 for SECCHI-B and 2007-present for SECCHI-A) CME catalogs, the Solar Eruptive Event Detection System (\href{http://helio.gmu.edu/seeds/}{SEEDS}), Automatic Recognition of Transient Events and Marseille Inventory from Synoptic maps (ARTEMIS) \citep{boursier2009automatic}, the CORIMP LASCO CME catalog \citep{byrne2012automatic}, and CIISCO \citep{patel2021automated}. These catalogs only include projected 2D properties. This is a major drawback to investigating the kinematic evolution of CMEs. For example, the projected speed and mass are, in general, a lower approximation of the true speed and mass. For many scientific and operational purposes, such 2D studies are inadequate  \citep{vrvsnak2004kinematics, zhang2004study,yashiro2003properties, gopalswamy2003coronal,gopalswamy2006coronal,yashiro2008comparison,gopalswamy2009expansion}. Various authors have aimed to overcome these effects by looking at CME properties estimated for events that occur close to the plane of the sky (i.e., over the solar limb), where projection effects are at their minimum \citep{michalek2003new,dal2003relation,michalek2009expansion,vasanth2013statistical}. 

CME observations from multiple viewpoints have the potential to give a more accurate and comprehensive understanding of CME kinematics. In the pre-STEREO era, CME reconstruction techniques like forward modeling, polarimetric, spectroscopic, and direct inversion have been extensively used to reconstruct 3D CME morphology. \cite{crifo1983coronal} initiated the direct reconstruction of CMEs, using polarization analysis to deduce that CMEs resembled 3D bubbles more than planar loops. Subsequent studies, such as \cite{jackson1995three}, using solar wind background models, confirmed this shape further in the heliosphere. This finding aligns with the observed 3-part CME structure, characterized by a bright front, a darker cavity, and a bright core, as described by \cite{illing1985observation}. By 1996, with the start of the SOHO mission, the earlier concept of planar, loop-like CMEs was largely replaced by understanding CMEs as 3D structures, particularly for events with this 3-part configuration. \cite{frazin2009toward} has determined CME density structures with only three viewpoints using image processing methods. \cite{thernisien2011cme} reviews the pre-STEREO efforts in 3D CME reconstruction and compares them with the post-STEREO results. In summary, the multi-viewpoint vantage provided by both SOHO and STEREO has prompted various techniques to investigate the 3D geometrical and kinematical information of CMEs \citep{chen1997evidence,wood1997lasco,michalek2003new,zhang2003identification,xie2004cone,schwenn2005association,gopalswamy2009expansion,mierla2008quick,moran2010three,feng2013comparisons,aschwanden20094,wood2009reconstructing,xie2009origin}. More recently, flux rope models have been utilized effectively and suggested by many authors \citep{chen2006cme, vourlidas2000large, cremades2004three} to generate CME morphology \citep{chen1997evidence} and to investigate their properties \citep{vourlidas2000large, krall2006all}. An effective tool for the 3D reconstruction of CMEs, the Graduated Cylindrical Shell (GCS) model, developed by \citet{thernisien2006modeling,thernisien2009forward} was originally implemented to simultaneously fit COR1 and COR2 images \citep{thernisien2011implementation} and later to LASCO-C2/C3 images \citep{shen2013full,shen2014full,colaninno2013quantitative}, as well as to the ground-based K-Cor \citep{majumdar2022variation} images. The model consists of a main structure with a tubular cross-section with a radius increasing with height and two cones attached to its ends, separated by twice the half angle $\alpha$. By fitting an observed CME image to the GCS model, we are assuming that the CME has an axisymmetrical flux-tube geometry that is expanding with distance from the Sun. This is a reasonable assumption that gives CME properties that are closer to the true CME properties compared to a simple 2D analysis. Recently, catalogs like the Coronal Mass Ejection Kinematic Database (KINCAT) and Coronal Mass Ejection Database (CMEDB) are based on 3D CME parameters estimated from using the GCS model.  

It is known that CME kinematics in the corona differ depending on their source location and eruption type \citep{hundhausen1993sizes}. CMEs associated with different source region types have been extensively studied and are broadly classified into active regions (ARs), prominences (PEs), and active prominences (APs) \citep{howard1986solar,munro1979association,subramanian2001source, moon2002statistical, majumdar2020connecting}. AR-CMEs are mostly impulsive and fast with a stronger magnetic field, whereas PE-CMEs are gradual and slow with a weaker magnetic field \citep{macqueen1983kinematics, sheeley1999continuous}. Historically, prominences have been classified based on their morphology and activity e.g., the
active, eruptive, sunspot, tornado, and quiescent prominences of \citet{pettit1925forms,mcmath1938prominence,pettit1943properties}. More recently, prominences have commonly been placed into two categories: active region and quiescent \citep{gilbert2000active}. AP CMEs are associated with prominences with either or both of their footpoints within an active region. \citet{pettit1925forms} describes active prominences as when material flows into nearby active regions, altering the magnetic topology of the constraining magnetic environment. The one exception to this categorization may be some stealth CMEs which are thought to form higher in the solar atmosphere \citep[e.g.][]{alzate2017}. The effect of various source locations on the width distribution of CMEs was recently reported by \cite{pant2021investigating}, who found that the widths of CMEs followed power law distributions, with some interesting differences in power law exponent according to the CME source region type. To determine whether or not the source regions have a measurable effect on CMEs, such big statistical studies are needed to examine the speed and width of CMEs in the outer corona based on their source region type. Unfortunately, \citet{pant2021investigating} was necessarily restricted to 2D projected widths only, although some mitigation was provided through the selection of limb events. Larger statistics of CMEs fitted to a 3D geometry are rare. \citet{jang2016comparison} used the Stereoscopic CME analysis tool (StereoCAT) to calculate 3D parameters, compared the 2D and 3D properties of 306 Halo CMEs, and found that 2D speeds underestimate 3D speeds by 20 \%. They give estimates of average 3D speed and average 3D width and compare 3D speed-width relationship to their corresponding 2D relation. Previous research has made valuable contributions to our understanding of CMEs and their kinematics. However, studies utilizing the GCS technique are limited to a smaller number of cases and do not investigate the relationships between CME speed, width, and different types of CME source regions from a comprehensive statistical perspective. This study addresses these gaps in the literature.

This work analyses 360 CMEs fitted using the GCS technique. The sample is separated into three groups according to their source regions to find statistical differences between the groups, and 3D (`true') and 2D projected CME parameters are compared. The datasets and methodology are described in sections \ref{sec:data} and \ref{sec:method}. Section \ref{sec:results} presents the primary findings with discussion, while Section \ref{sec:conclusion} offers summary and conclusions.

\section{Data Anaylsis}
    \label{sec:data}
\subsection{Data Selection and Preparation}
First, we present a description of the chosen datasets and a justification for the decisions made throughout this research. We then go on to detail the subsequent preprocessing methods for the data. Note that in this section and elsewhere, that the 3D fitted CME parameters are referred to as `true', although this is, of course, subject to the assumption that the GCS geometrical model is the correct choice of geometry for all CMEs.

Our comprehensive database of 360 CMEs includes the original CME data (cdaw date, first C2 appearance time, position angle, projected speed, and projected width from automated and manual methods) and the corrected speed, width, and source region type from 2007-2021, covering the rising and declining phase of solar cycle 24. There is a data gap from October 2014 to November 2015 when STEREO A went behind the Sun. We select only those CMEs with a clear three-part morphology, which appear bright and well-structured in the SECCHI/COR2 field of view. In line with this, we have combed through archival papers and catalogs to compile a list of CME features that meet the aforementioned criteria. Given the manual and subjective nature of fitting the GCS model to CMEs, we developed a selection criterion for CME events to reduce subjective biases. Our approach focused on including events already documented in existing catalogs, such as KINCAT and CMEDB, and those discussed in research papers where the GCS model fitting was systematically applied to multiple CME events. In contrast, we deliberately omitted isolated case studies like the September 2017 event, which is extensively covered in various studies (e.g., \citep{gopalswamy2018extreme,shen2018shock,werner2019modeling,wu201904,scolini2019investigating,scolini2020cme}) to maintain consistency and minimize individual biases in our analysis. This was complemented by adding these events to our own dataset. As listed in Table \ref{table:data}, information on most of the 360 CMEs in this study is gained from the KINCAT and CMEDB. Other CMEs are extracted from published papers, as listed in Table \ref{table:data}, and 29 from the Data Research and More in Space Physics (DREAMS) catalog \citep{shen2013full,shen2014full}. A further 34 other CMEs are fitted by ourselves, as described in section \ref{fitting}. These 34 CMEs were not available in the other existing catalogs.  There are no duplicate events in the set of 360 CMEs in this study. The kinematic properties of the CMEs, such as speed and angular width, were directly available and recorded for our study.

\begin{table}[h]
\renewcommand{\arraystretch}{1.3}
\resizebox{\linewidth}{!}{%
\centering
\begin{tabular}{l|l|l|l}
\hline
  \textbf{No.} &  \textbf{No. of CMEs} &  \textbf{3D Catalogs/past studies} &  \textbf{2D Catalogs used for comparison} \\
  \hline
  1 & 119 & CME Kinematic Database \href{http://www.affects-fp7.eu/helcats-database/database.php}{KINCAT} & \textcolor{black}{Automated catalogs [\href{https://www.sidc.be/cactus/}{CACTus} (No. of events = 342)} \\
  2 & 79 &  \href{http://www.affects-fp7.eu/cme-database/database.php}{CME Database} & \textcolor{black}{and \href{http://spaceweather.gmu.edu/seeds/}{SEEDs} (No. of events = 18)}  \\
  3 & 50 & \citet{majumdar2020connecting}  & \textcolor{black}{from STEREO-A viewpoint ]} \\
  4 & 07 & \citet{cremades2020asymmetric}  \\
  5 & 29  & \href{http://space.ustc.edu.cn/dreams/fhcmes/}{DREAMS} & \textcolor{black}{and} \\
  6 & 23 & \citet{pal2018dependence} \\
 7 & 19 & \cite{lee20153} & \textcolor{black}{Manual Catalog [\href{https://cdaw.gsfc.nasa.gov/CME_list/index.html}{CDAW} (No. of events = 355)}  \\
  8 & 34 & This study & \textcolor{black}{from LASCO viewpoint]}  \\
\end{tabular}}
  \caption{Collection of CMEs and associated catalogs, containing 34 CMEs fitted during this study and 360 CMEs studied in total.}
  \label{table:data}
\end{table}

The definition of parameters relating to CME size (or angular width) differs between the various studies and catalogs, however, we were able to use the parameters such as half-angle and aspect ratio to determine a consistent angular width. The GCS fitting procedure has been conducted on 360 events using coronagraph images from two viewpoints, namely STEREO-A and STEREO-B for events from 2007 until 2014 October, and STEREO-A and LASCO for events occurring after 2015. Our selection criteria for 2D catalogs aimed to ensure that the catalogs contained detections in one of the coronagraph images: LASCO, STEREO-A, STEREO-B, or all three. Automated catalog CACTus detects CMEs in STEREO and LASCO whereas manual catalog like CDAW only detects CMEs in LASCO. We made a list of  STEREO/SECCHI-COR2 projected parameters from the widely used automatic catalogs CACTus and SEEDS and LASCO projected parameters from CDAW while keeping in mind that any catalog, whether automatic or manual, would have computational and human biases, with detections affected by projection effects \citep{yashiro2008comparison} . Hence, by using pre-existing 2D CME databases, we conducted a statistical analysis to compare the actual kinematic qualities with the expected kinematic features.  

\begin{figure}[h]
\centering
 \includegraphics[width=0.9\textwidth]{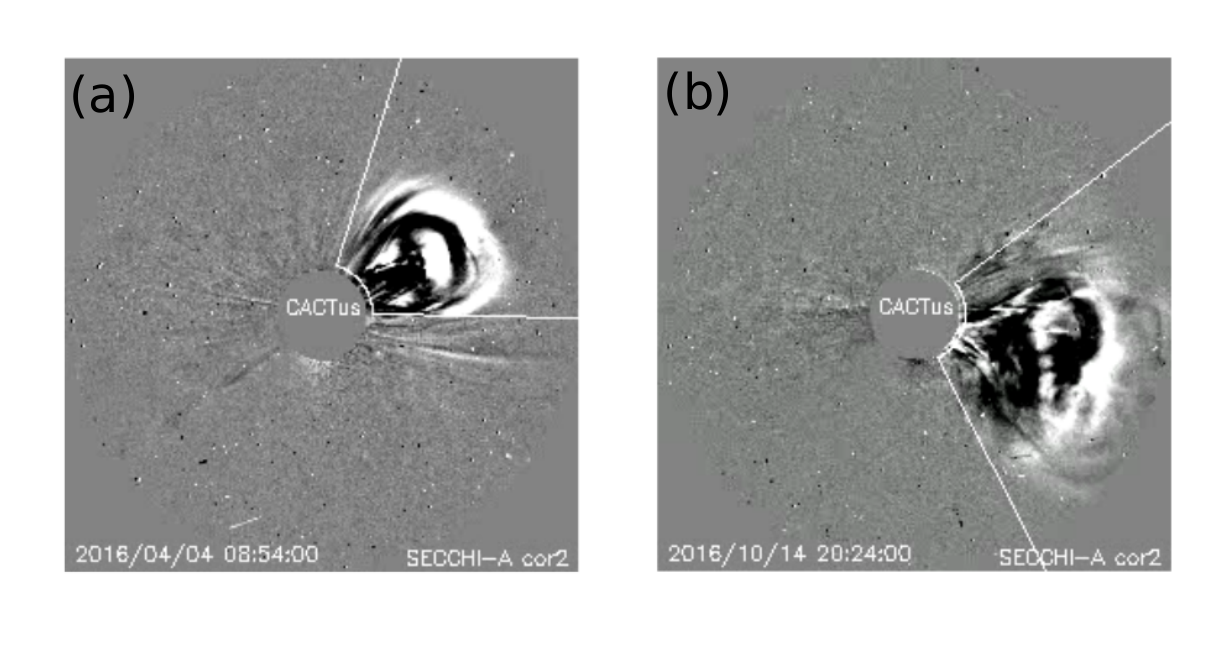}
   \caption{(a) CME of 2016/04/04 at 08:54 UT fiited by CACTus in COR2A (b) CME of 2016/10/14 at 20:24 UT fitted by CACTus in COR2A.}
\label{fig:cactus}
\end{figure}

CACTus was the reference catalog for projected parameters for 342 events, whereas 18 events not cataloged in CACTus were taken from SEEDS. However, both CACTus \citep{robbrecht2009automated} and SEEDS \citep{olmedo2009automatic} are automated catalogs; they detect and track the front of CMEs in running difference images differently to provide an estimation of the CME projected parameters. \cite{hess2017comparing} compares similarities and differences between SEEDS and CACTus and shows a strong correlation between these detection techniques and datasets when comparing them across LASCO and SECCHI. When we compare estimated speeds calculated automatically using CACTus or SEEDs to those derived by visual observation, we find substantial differences \citep{braga2013pseudo}. Sometimes, CMEs can drive shocks in the solar wind and can have significant implications when cataloging and measuring the properties of CMEs \citep{kahler2009cme}, especially the angular width. In instances where a shock front is closely aligned with a CME, the effective width observed may indeed be an amalgamation of both features, and the brightness enhancement could be seen extending beyond the actual boundary of the CME ejecta. Automated algorithms like CACTus identify CMEs by detecting transient changes in brightness in coronagraph images (Figure \ref{fig:cactus} shows two CMEs fitted with CACTus).
Consequently, CACTus may overestimate the width of the CME by including the shock in its measurement. On the other hand, manual fitting methods like the GCS model focus on capturing the three-dimensional structure of the CME by focusing on the core of the CME with a flux rope structure, potentially neglecting the shock structure. As a result, the GCS model's width may underestimate the event's total spatial extent as observed in a coronagraph. Some bias will always be involved regardless of the fitting methods, such as subjectivity from human inspections or misjudgment based on image processing thresholds in automated methods. Hence, both automated and manual methods can be affected by the presence of shocks associated with CMEs, leading to potential inaccuracies in the determined widths. Hence, in this context, using either of the projected sets of parameters is valid as both methods yielded similar correlations for speed-width distributions.
\subsection{Source region determination}
\begin{figure}[h]
    \centering
    \includegraphics[width = 0.8\textwidth]{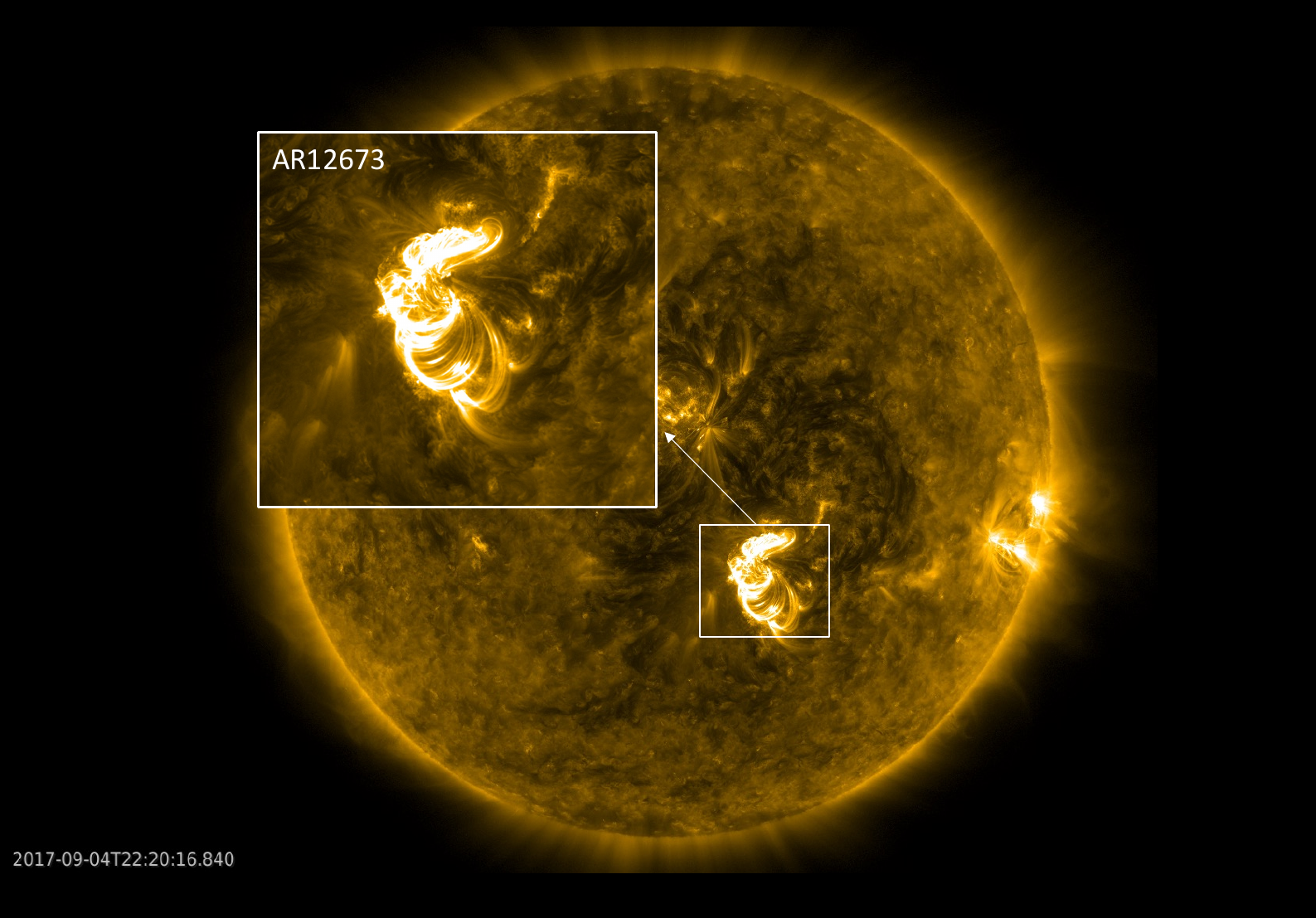}
        \caption{A source region AR12673 that erupted 2017/09/04 CME is shown in AIA 171A with post eruptive loops.}
        \label{fig:AIA171}
\end{figure}

\begin{figure}[h]
    \centering
    \includegraphics[width = 0.8\textwidth]{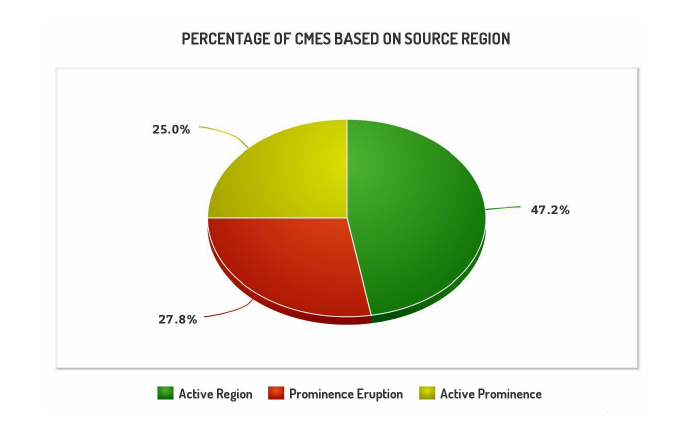}
        \caption{CME's source region classification into types active region, active prominence, and prominence eruption.}
        \label{fig:piechart}
\end{figure}

CMEs were divided into three categories based on their source region type:  active regions, prominences, and active prominences. Data from the Solar Dynamics Observatory's Atmospheric Imaging Assembly (AIA; \cite{lemen2012atmospheric}), the Extreme Ultraviolet Imaging Telescope (EIT; \cite{delaboudiniere1995eit}) on board SOHO and SECCHI EUVI were used to identify the source locations of CMEs emerging from the Sun's surface. JHelioviewer \url{https://www.jhelioviewer.org/}  \citep{muller2009jhelioviewer,muller2017jhelioviewer}, a visualization software for solar image data was used to access the data. To identify the source region (SR) for frontside CMEs in the low corona, EIT observations in 195 \AA\ and 304 \AA\ from 2007 to 2009, and AIA observations in the 171\AA\, 193 \AA\ and 304\AA\ channels post-2009 were examined for pre- and post-eruptive signatures of CMEs, such as the outward motion of coronal material in the form of loops (Figure \ref{fig:AIA171} shows loops after CME eruption from AR12673) or eruptive prominences \citep{webb1987activity} and post-eruptive loops/arcades \citep{sterling2000yohkoh, cremades2004three} whereas SECCHI EUVI 195 \AA\ and 304 \AA\  channels were used to identify source region signatures for the backside CMEs with respect to the Earth.
For a comprehensive explanation of the identification process for source regions, see \cite{majumdar2020connecting} and \citep{majumdar2023cme}. As shown in Figure \ref{fig:piechart}, 170 (47.2\%) of the cataloged events fall into the AR category, 90 (25\%) fall into the AP category, and 100 (27.8\%) fall into the PE category. 

\section{Methodology}
\label{sec:method}
\subsection{3D Reconstruction and Estimation of Geometric Parameters}
\label{fitting}
\begin{figure}
    \centering
   
        \includegraphics[width=\textwidth]{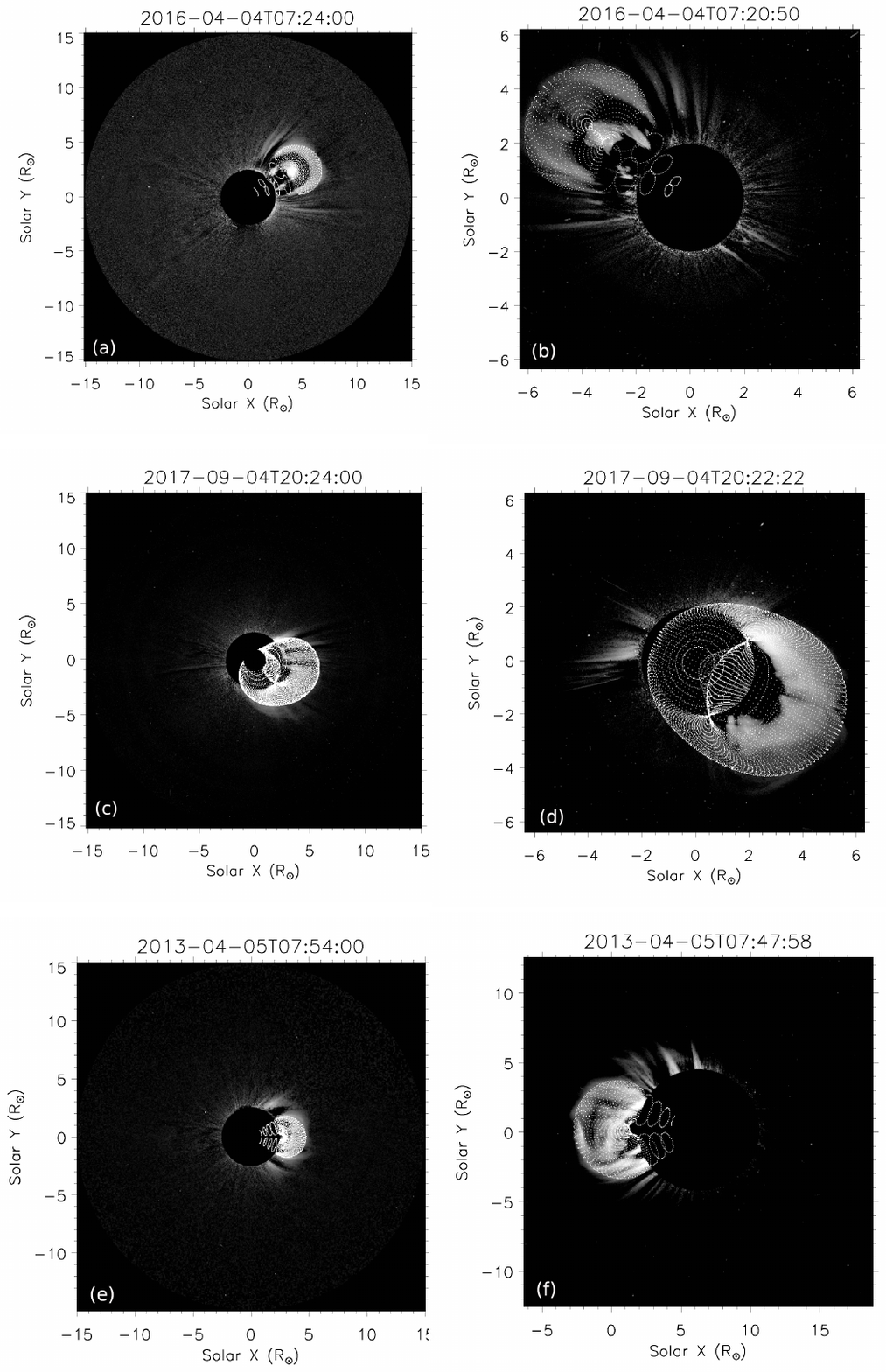}
    \caption{Fittings of the GCS flux rope to the COR-2A (left) and C2 (right) images for three CMEs. Panels (a) and (b): CME of 2016/04/04 at 06:24 UT. Panels (c) and (d): Halo CME of 2017/09/04 at 20:24 UT. Panels (e) and (f): CME of 2013/04/05 at 07:47UT}
    \label{fig:GCSincor2c2}
\end{figure}

The 34 CMEs we fitted using the GCS model were chosen to improve the statistical sample size. The data used for this sample are taken from the STEREO COR-2A coronagraphs, the STEREO Extreme UltraViolet Imager (EUVI), and LASCO/SOHO C2/C3. The level 0.5 data of EUVI and COR-2 were reduced to level 1.0 using the secchi\_prep.pro routine in the Solarsoft library of the Interactive Data Language (IDL), and for LASCO, we used level 1 data (corrected for instrumental effects and solar north and calibrated to physical units of brightness) \citep{majumdar2020connecting}. Further, they were processed with the Simple Radial Gradient Filter \citep[SiRGraF:][]{Patel2022} utilizing the full daily images to create the necessary backgrounds. The processed SiRGraF images were then used to fit the CMEs to the GCS geometrical model. To capture evolution in the outer corona, the model is fitted simultaneously to the COR-2 (FOV of 2.5–15 $ R_{\odot}$) and C2/C3 (FOV of 2.2–30 $ R_{\odot}$) images. The GCS fitting for two limb CMEs at 2016/04/04 06:24 UT and 2013/03/05 07:54 UT and a halo CME at 2017/09/04 20:24 UT is shown in figure \ref{fig:GCSincor2c2}. As a result, fitting parameters (see Table \ref{table:GCSparams}) such as height, aspect ratio, half-angle, tilt-angle, latitude, and longitude were recorded. Since three–vantage point observations are not available since STEREO-B stopped functioning in 2014, some model parameters (Latitude, Longitude, tilt-angle) were fixed according to their source region location, while the height, half-angle, and aspect ratio were fitted to the time series of images. The fitting procedure was carried out according to the descriptions of \citet{thernisien2006modeling, thernisien2009forward} and \citet{majumdar2020connecting}. The fitted geometrical parameters are listed in Table \ref{table:GCSparams} with CME speed and width derived from GCS and projected parameters taken from automated and manual catalogs for comparison.

\newpage
\subsection{Estimation of 3D Kinematic properties}
\begin{figure}[h]  
\centering
 
   \includegraphics[width=\textwidth]{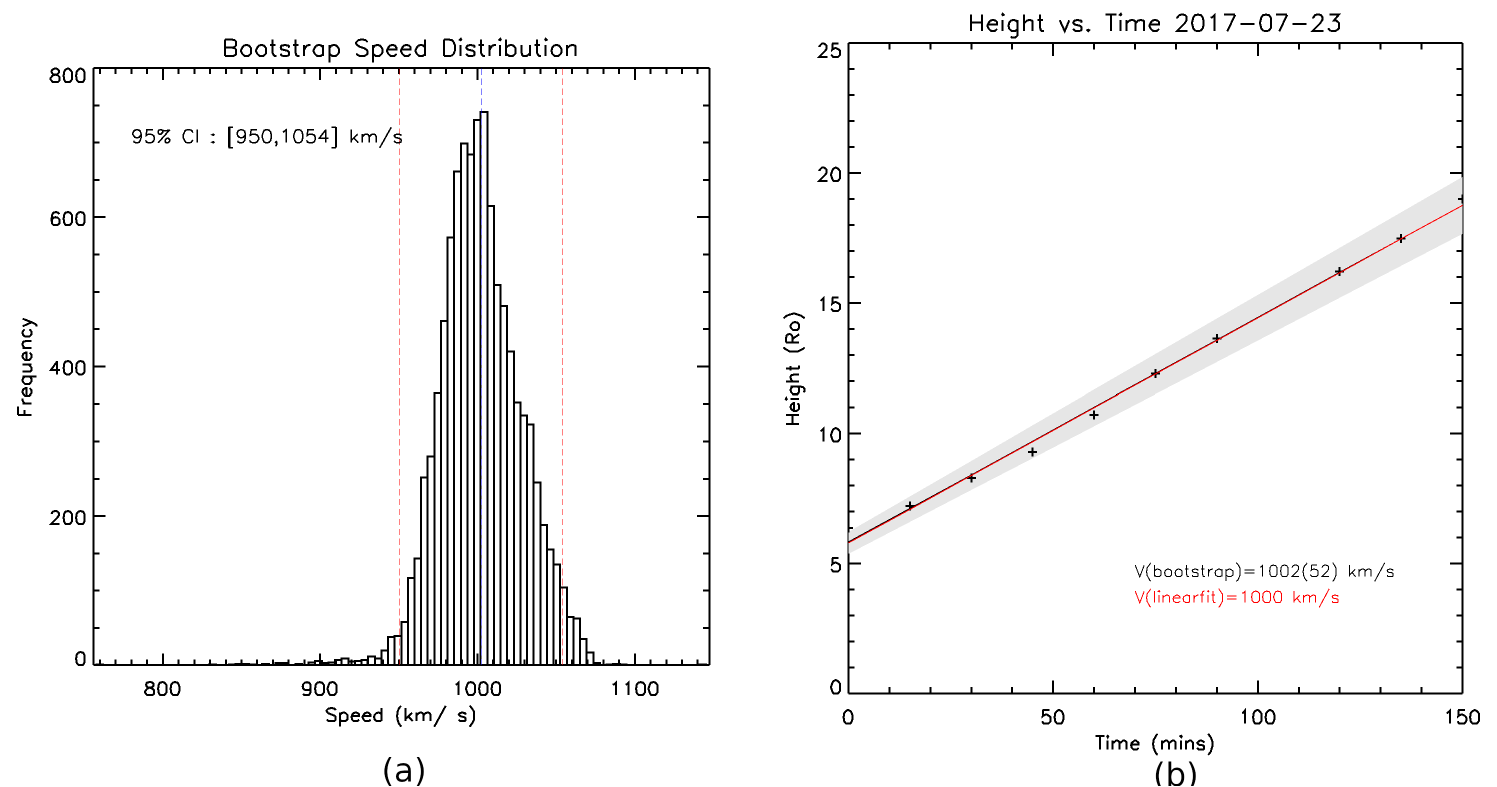}
 
 \caption{(a) Distribution of speed estimates arising from a bootstrapping resampling method (see text), with 95\%\ confidence intervals (CI) between the dotted red lines. (b) An example of a height-time plot for a CME occurred at 01:25 UT on 2017-07-23. The red line shows the linear fit, and the shaded area indicates the 95\%\ uncertainty arising from the bootstrap distribution.}
\label{fig:bootstrap}
\end{figure}

Following the GCS geometrical fitting, the average speed of the 34 CMEs is estimated using height-time information, and as shown in \ref{fig:bootstrap}b, linear regression is used to fit the data. To maintain consistency in the speed estimates for all 360 CMEs, a simple linear relationship of distance to time is maintained. Since GCS is a manual technique for fitting CMEs, there is a possibility of human bias being included in the parameters that are modeled. Using a range of possible values for each fitting parameter helps to confine better the estimations like the speed and width of the CME. This could be done by performing the fitting procedure multiple times to lower the degree of uncertainty and increase confidence in the fitting parameters or by employing a resampling method instead to carry out the same tasks. Keeping this in mind, we used a resampling technique called bootstrapping. First introduced by \citet{efron1986bootstrap} and more recently described in \citet{efron1994introduction,chernick1988bootstrap} and \citet{byrne2013improved}, the method works by resampling the original dataset thousands of times with randomly-ordered residuals to generate simulated datasets. In the case of CMEs, bootstrapping techniques can be beneficial for estimating the uncertainty in speed. Here, we have utilized the IDL Boot$ $XYFIT procedure on height and time measurements, which aims to perform a linear fit to the data with errors in both $X$ (time) and $Y$ (height) where measurement errors are unavailable. A histogram of the speed sample estimated by BOOT\_XYFIT is shown in figure \ref{fig:bootstrap}a, where the 95\%\ confidence interval is shown as dotted red lines. This range defines the speed error, shown as the grey shaded region in figure \ref{fig:bootstrap}b. Since we do not know the exact kinematic form a CME should take nor the true uncertainty due to possible unknown sources of error, a bootstrapping technique allows an appropriate confidence interval to be assigned to the kinematic parameters. Table \ref{table:GCSparams} lists the GCS fitting parameters along with GCS speed and width for all 34 CMEs.
\section{Results and Discussion}
    \label{sec:results}
Table \ref{table:GCSparams} summarises the parameters for the 34 CMEs fitted during this study. Listed in columns 1 through 6 is the total number of CME occurrences, the dates and times when CMEs were first detected in LASCO FOV, their SR type, and the latitudes and longitudes of the SR locations. The other columns list the GCS parameters. True width ($W_{GCS}$) in column 10 is calculated using the formula (2$\alpha$+2$\delta$) where $\delta$  = $\arcsin$($\kappa$) and $\alpha$ and $\kappa$ are the GCS half-angle and aspect-ratio shown in columns 8 and 9. Columns 10 and 11 show true width and speed estimated from GCS with uncertainty calculated using Bootstrap. Projected width and projected speed from automated and manual methods are listed in columns 10 to 15.

\begin{table}[h]
 \renewcommand{\arraystretch}{1.25}
    \resizebox{\linewidth}{!}{
 \begin{tabular}{c c c c c  c c c c c c c c c c} 
 \hline 
 
 \textbf{No.} & \textbf{Date} & \textbf{Time [UT]} & \textbf{SR}  & \textbf{Lat($\theta)$} & \textbf{Long($\phi)$} & \textbf{Tilt($\gamma$)} & \textbf{aspect-ratio($\kappa$)} & \textbf{half-angle($\alpha$)} & \textbf{$W_{GCS}$($^{\circ}$)} &  \textbf{$V_{GCS}$(km$^{-1}$s)} & \textbf{$V_{CACTus}$(km$^{-1}$s)} & \textbf{$W_{CACTus}$($^{\circ}$)} & \textbf{$V_{CDAW}$(km$^{-1}$s)} & \textbf{$W_{CDAW}$($^{\circ}$)}\\
 \hline
1 & 2013-04-05 & 06:36 & PE & 2 & 243 & 65 & 0.24 & 30 & {88} & 575(15)&543&137&588&228\\
2 & 2013-04-08 & 08:36 & PE &  -28 & 120 & -45 & 0.23 & 10 & 52& 555(18)&413&52&529&85\\
3 & 2013-04-08 & 13:25 & AR & -17 & 120 & 50 & 0.20 & 10& 42& 395(12)&427&55&393&89\\
4 & 2016-01-01 & 23:24 & AP & -26& 320 & 50& 0.25 & 20 & 69& 1580(25)&1562&328&1730&360\\
5 & 2016-01-04 & 23:36 & PE & 54 & 279 & -12 & 0.25& 10 &78& 236(21)&431&62&250&164\\
6 &2016-01-06 & 14:00 & PE & -19 & 284 & 45  & 0.24 & 12& 52 & 370(13)&353&63&969&360\\
7 & 2016-01-29 & 21:15 & AR	& -22 & 110 & -27 & 0.32 & 18& 73& 925(53)&657&76&901&118\\
8 & 2016-02-20 & 14:24& AP & 6 & 139& 90 & 0.36 & 26 & 94& 660(20)&446&170&491&360\\
9 & 2016-04-04 & 03:24 & PE	& 10& 232 & 60  & 0.22& 12 & 50& 518(47)&446&74&444&129\\
10 & 2016-05-10 & 23:36 &	AR &	-36	& 268 & 29 & 0.23 & 10 & 45 & 368(30)&347&76&327&36\\
11 &  2016-05-15	& 15:12	& 	AR &	15 &	60 & 54 & 0.56 & 14& 96 & 962(45)&892&146&1118&176\\
12 & 2016-06-25	& 02:48	&	AP	& 20	& 285 & 67 & 0.32& 14 & 65& 489(17)&378&96&540&128\\
13 & 2016-09-06 & 08:36	&	AP &	-20 & 	114 & 33 & 0.25 & 33 & 95 & 688(44)&694&70&650&101\\
14 & 2016-10-14 & 09:48	&	AP & -20& 270 & 37 & 0.37 & 12 & 67 & 511(20)&510&100&375&75\\
15 & 2016-10-14 & 16:48	& AR& -15& 273 & 42 & 0.29 & 12& 58& 646(52)&510&100&540&104\\
16 & 2016-12-12 & 06:48	&	PE & 28 &	151 & 25 & 0.22 & 13 & 51& 622(78)&595&52&424&152\\
17 & 2016-12-21 &	20:00 &		AR &	1.6 & 302 & -69 & 0.28 & 12 & 57& 202(12)&211&58&207&93\\
18 & 2017-03-13	& 20:12 &	AP &15	& 140 & -61 & 0.21 & 22 & 69& 465(35)&341&106&668&221\\
19 & 2017-04-18	& 19:48 &	AR & 18	& 285 & 70 & 0.25  & 30 & 88& 906(46)&892&154&926&360\\
20 & 2017-07-14	& 01:25	&	AR &-10.6 &	35.8 & -31 & 0.68 & 41& 167& 1460(25)&892&202&1200&360\\
21 & 2017-07-17	& 16:36	&	AR &	-6 &	71 & 83 & 0.19 & 25 & 72 & 512(37)&337&90&376&122\\
22 & 2017-07-20	& 18:12	&	AR	& 1	& 125 & 58 & 0.30 & 22 & 79& 686(63)&500&82&590&95\\
23 & 2017-07-23	& 01:25 &	AR & -5 & 	146 & -52 & 0.70 & 35 & 66 & 1002(52&756&190&654&144)\\
24 & 2017-07-23	& 04:48 &	AR &	-5 & 146 & 63 & 0.19 & 22 & 160 & 2039(569)&1388&352&1848&360\\
25 & 2017-09-04	& 20:36 &	AR	& -10	& 12 & -24 & 0.48 & 18 & 92 & 997(99)&462&274&1418&360\\
26 & 2017-09-06	& 12:24	&	AR & -8 & 	36 & 26 & 0.65 & 19 & 119  & 1591(114)&735&352&1571&360\\
27 & 2017-09-09	& 16:24	&	AR &	-8 & 105 & 45& 0.24 & 18 & 64 & 609(144)&403&72&473&95\\
28 & 2017-09-10	& 16:00	&	AR &	-11 &	91 & 80 & 0.54 & 35 & 135 & 2320(150)&1785&352&3163&360\\
29 & 2017-09-17	& 12:00 &	AP & -6 &	189 & -39& 0.55 & 22 & 110& 1910(250)&1250&166&1385&360\\
30& 2017-10-18&05:48&AP&-13&260& -52 & 0.61 & 17 & 109 & 1530(163)&600&126&1576&360\\
31 & 2019-04-30 & 14:48 & AP & -5 & 192 & 17 & 0.43 &37 & 125 & 1040(85)&781&94&665&360\\
32 & 2020-12-07 & 16:24 & AP & -22 & 7 & 58 & 0.41& 24 & 95 &1560(45)&584&359&1407&360\\
33 & 2021-10-09 & 07:12 & AR & 19&5 & 45& 0.32& 22 & 81&1589(33)&781&152&712&360\\
34 & 2021-10-28 & 15:48 & AR & -27 & 1 & 8 & 0.60 & 40 & 153 & 1369(23) & 924&121&1519&360\\
\hline
\end{tabular}}
\caption{The GCS Model Parameters of the 34 CMEs that were fitted during this work, together with their true width (column 9), average true speed (column 10), and speed calculated via bootstrap with errors (column 11).}
\label{table:GCSparams}
\end{table}

 A summary of average CME properties with and without SR segregation, before and after correcting projection effects, is provided in Table \ref{table:AVGCME}. This table presents a comprehensive overview of the observed properties of CMEs classified by different source regions. It includes aggregated statistical parameters for all CMEs under study (360 events) and subsets categorized by source regions: '170 AR' (Active Region), '90  AP' (Average Prominence), and '100 PE' (Prominence Eruption) CMEs. For each category, the table lists the mean and median values of the 3D Angular Width, 3D Speed, 2D Angular Width, and 2D Speed. Additionally, the slope and associated error of the speed-width relation, along with the 95\% confidence intervals for both the 3D and 2D slopes, are provided. These intervals offer insights into the reliability and variability of the slope estimates. The data indicate notable differences across categories, reflecting the inherent diversity in CME dynamics. \citet{howard2008three} shows a similar table with the mean in average apparent and corrected speeds.
 
\begin{table}
 \renewcommand{\arraystretch}{1.5}
    \resizebox{\linewidth}{!}{%
 \begin{tabular}{l l l l l} 
 \hline 
 CME Property & ALL (360) CMEs & 170 "AR" CMEs & 90 "AP" CMEs & 100 "PE" CMEs\\
 \hline
 GCS Angular Width (mean, median in $^{\circ}$) & 77, 72 & 79, 74 & 85, 80 & 66, 62 \\
 GCS Speed (mean, median in km s$^{-1}$) & 665, 561 & 723, 642 & 813, 690 & 432, 388\\
 CACTus Angular Width (mean, median in $^{\circ}$) & 112, 93& 122, 100 & 133, 117 & 75, 69\\
 CACTus Speed (mean, median in km s$^{-1}$) & 499, 446 & 534, 491 & 592, 531 & 358, 337 \\
 CDAW Angular Width(mean, median in $^{\circ}$) & 189, 153 & 202, 169 & 234, 221 & 123, 109\\
 CDAW Speed(mean, median in km s$^{-1}$) & 613, 527 & 656, 553 &782, 650 &390, 347  \\
 slope, error in slope and cc of GCS speed-width & 5.81, 0.63, 0.44 & 7.80, 0.97, 0.52& 4.57, 1.20, 0.37 & -0.78, 0.71, -0.10\\ 
 95\% confidence interval for GCS slope & (4.57, 7.05) & (5.88, 9.72)& (2.18, 6.96) & (-2.20, 0.63) \\
 slope, error in slope and cc of CACTus speed-width & 2.19, 0.15, 0.61 & 2.07, 0.20, 0.62 & 2.07, 0.20, 0.50&1.47, 0.37, 0.36 \\ 
 95\% confidence interval in slope & (1.89, 2.49) & (1.67, 2.47) & (1.23, 2.63) & (0.73, 2.23) \\
slope, error in slope and cc of CDAW speed-width & 2.14, 0.14, 0.60 &2.27, 0.23, 0.59&1.83, 0.34, 0.49&0.97, 0.19, 0.45 \\
 95\% confidence interval in slope & (1.85, 2.42)&(1.80, 2.73)&(1.15, 2.51)&(0.58,1.37)\\
\hline
\end{tabular}}
\caption{Shows the statistical parameters, including mean and median values for angular width and speed, slope errors, and 95\% confidence intervals for the slope estimates, segregated into overall, AR-CMEs, AP-CMEs, and PE-CMEs groups before and after correction.}
\label{table:AVGCME}
\end{table}
\newpage
\subsection{Comparing Speed and Width Distributions before and after correction}
\begin{figure}[h]  
\centering
    \includegraphics[width=\textwidth]{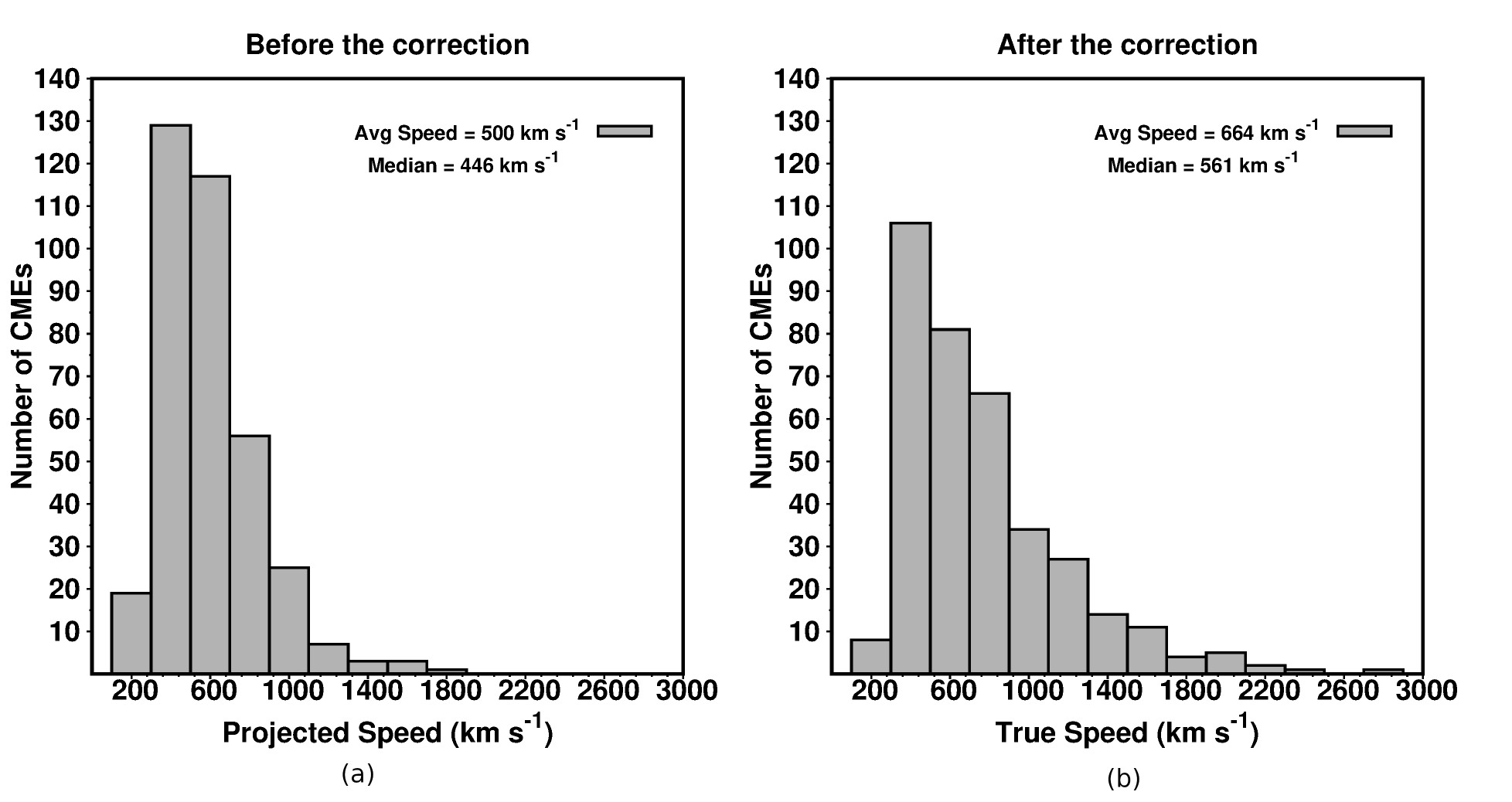}
\caption{Speed distributions of 360 CMEs, represented by vertical bars, before correction \textit{(left panel)}, and after correction using the GCS approach \textit{(right panel)}, together with the average speed of the CMEs. The bin size is 200  km $s^{-1}$.}
\label{fig:2D3DSD}
\end{figure}

Figure \ref{fig:2D3DSD} shows the distributions of apparent speeds (2D) of CMEs (left panel) and the true speeds (3D) of CMEs using GCS (right panel). The distributions are not symmetrical, with the maximum number of CMEs with speeds peaking around 300-500  km s$^{-1}$ in both cases as expected. The range of true radial speeds (figure b) is much higher (100-2900  km s$^{-1}$) than the projected speeds (figure a), ranging from 100-1900  km s$^{-1}$. The figure demonstrates that after correcting the projection effects, the high-speed tail of the distribution is significantly boosted. In contrast, the low-speed tail is significantly suppressed, with the number of CMEs with speeds between 100-300  km s$^{-1}$\ reduced by half. The average speed of a CME is 500  km s$^{-1}$ before the correction; however, the GCS approach produces an average speed of 665 km s$^{-1}$\, showing that 2D speed tends to be about 30\% underestimated compared to the 3D ones. The 3D distribution has a higher median speed (561 km s$^{-1}$) compared to the 2D distribution (446 km s$^{-1}$), indicating that the corrected speeds tend to be higher overall. 

There is a more significant spread of speeds in the 3D distribution, as evidenced by a larger standard deviation (415 km s$^{-1}$ vs. 255 km s$^{-1}$) and a wider interquartile range (493 km s$^{-1}$ vs. 306 km s$^{-1}$). This implies that the corrected data have a broader range of values and potentially more variability or dispersion around the median.
A large number of studies [for example, \citet{burkepile2004role,vrvsnak2007projection,howard2008kinematic,howard2008three,yeh2005kinetic,xie2009origin,shen2013full,jang2016comparison}] have looked at the distribution of kinematics of CMEs, with and without projection effects. \citet{howard2008kinematic} compared apparent and true speed histograms from 1996 - 2005 on a yearly basis and showed that the magnitude of corrected measurements differs significantly from the projected plane-of-sky measurements with average speeds ranging from 625-900  km $s^{-1}$\ (true) and 270-435  km $s^{-1}$\ (apparent) which closely matches the average speeds in our study. \citet{shen2013full} showed using GCS that the true speeds of halo CMEs vary from 274  km $s^{-1}$\ to 2016  km $s^{-1}$\ with an average speed of 985  km $s^{-1}$.  The most extensive study to date, conducted by \citet{jang2016comparison}, used StereoCAT to estimate 3D parameters of 306 front-side halo CMEs during the rising phase of the solar cycle 24 and demonstrated that the 2D speed underestimates the 3D speed by approximately 20\%. Their result shows that the average speed of halo CMEs changes from 733 km $s^{-1}$\ to 896 km $s^{-1}$\ after correcting for projection effects. Prior studies have drawn similar conclusions when comparing 2D speeds with 3D speeds using methods such as triangulation [\citep{pizzo2004geometric};\citep{mierla2008quick};\citep{temmer2009cme}], STEREO-CAT [\url{http://ccmc.gsfc.nasa.gov/analysis/stereo/manual.pdf}], etc. Therefore, the results presented in this study not only support prior findings but also validate the effectiveness of various methodologies employed in reconstructing CMEs using multi-viewpoint images.

\begin{figure}[h]  
\centering

   \includegraphics[width=\textwidth]{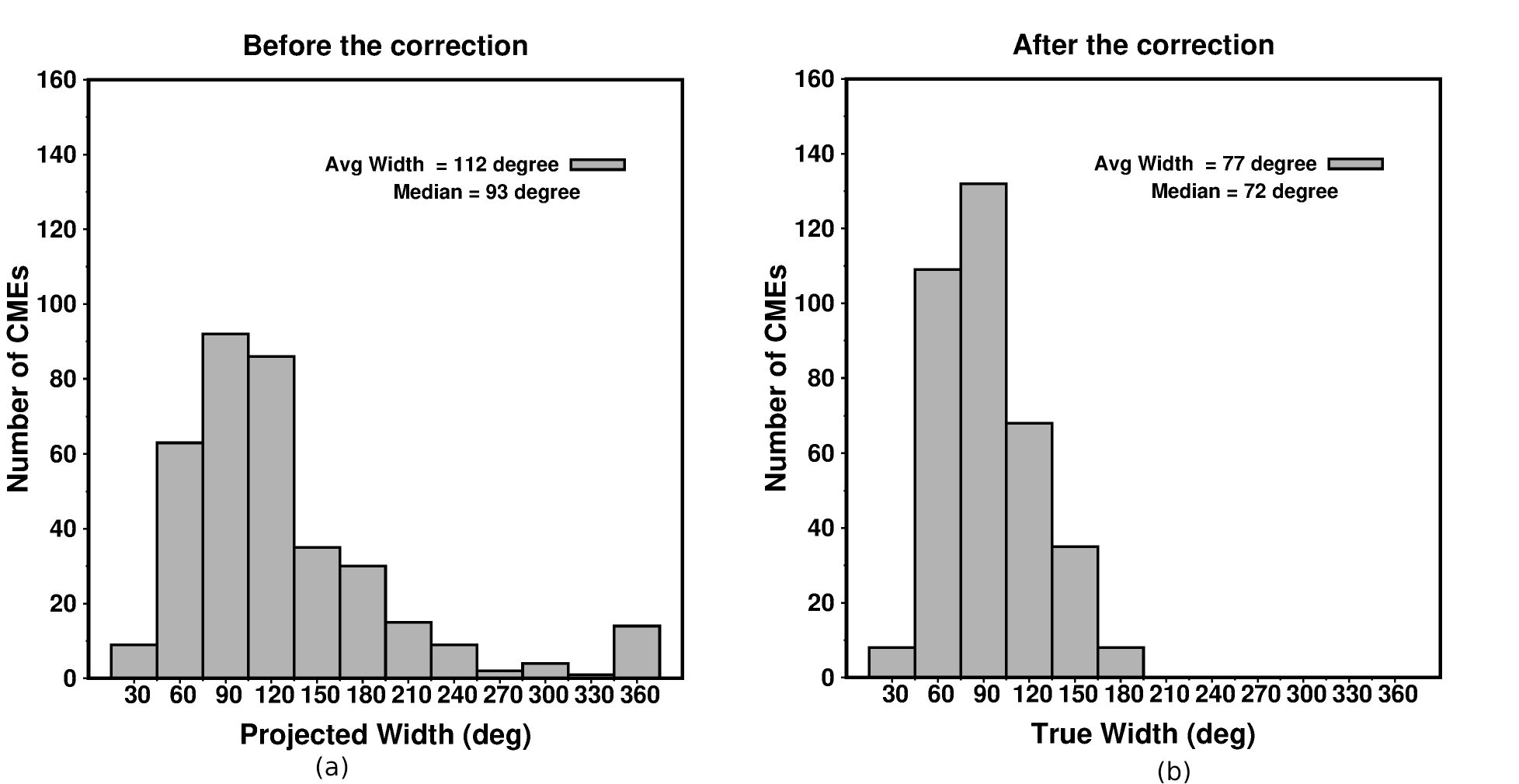}
  
\caption{Width distributions of 360 CMEs, represented by vertical bars, before correction \textit{(left panel)}, and after correction using the GCS approach \textit{(right panel)}, together with the average width of the CMEs. The bin size is 30 $^{\circ}$.}
\label{fig:2D3DWD}
\end{figure} 

The relationship between the apparent and true angular width of a CME, in general, is complicated \citep{st2000properties,yashiro2004catalog,howard2008kinematic,webb2012coronal,pant2021investigating}. The apparent angular width of a CME measured in the projected plane tends to be broader or narrower than the actual one. This could be due to the orientation and appearance (viewed edge-on or face-on) of CME in coronagraph FOV and the central longitude/latitude location of CME relative to the observer. The frequency distribution of the projected and true width is shown in Figure. \ref{fig:2D3DWD} as histograms with bins of 30$^{\circ}$, showing how the CME angular width spreads out before (left panel) and after (right panel) the correction. It can be seen that following the correction, the number of CME events that have wide angular spans (apparent width $>$ 180$^{\circ}$) drops to zero, while the number of events that have narrow angular spans shows a significant increase. The maximum frequency lies in the range 75$^{\circ}$ to 105$^{\circ}$ in both histograms, with an average true width of 77$^{\circ}$ and the average projected width  of 112$^{\circ}$. Studies that have compared 2D widths to 3D widths using techniques like triangulation and STEREO-CAT have also reached similar conclusions. These results not only corroborate those of previous research, but they also demonstrate the validity and difficulties associated with using different approaches to estimating the width of CMEs.

The projected/apparent average angular widths observed in previous studies vary widely, for example, the average width detected by the Solwind instrument between 1979 and 1981 was 45$^{\circ}$, and 24$^{\circ}$ between 1984 and 1985 \citep{howard1985coronal,sheeley1986solwind}. The average width for the 1300 CMEs recorded by the SMM instrument between 1984-1989 was 47$^{\circ}$ \citep{hundhausen1993sizes}, while the average width for the 240 CMEs recorded by the ground-based coronagraph MK3 between 1980 and 1989 was 37$^{\circ}$. \citep{st1999comparison}. Also, \citet{st2000properties} found that the average apparent width for 841 LASCO CMEs between January 1996 and June 1998 was 72$^{\circ}$.\citet{yashiro2003properties} calculated the yearly averaged width for LASCO CMEs from 1996 to 2001, and the results varied from 51$^{\circ}$ to 66$^{\circ}$. \citet{yeh2005kinetic} shows that the average width of a CME changes from 77$^{\circ}$ (apparent) to 58$^{\circ}$(corrected). \citet{jang2016comparison} found that the average width of partial and full halo CMEs changes from 227$^{\circ}$ to 83$^{\circ}$ after correcting the projection effects. 

Compared to the values above, we find a higher apparent average angular width in our case which overestimates the true average width by 30\%. In particular, the distribution of CME widths is considerably narrower than that of apparent widths. The overall shape of the distribution seen in Figure \ref{fig:2D3DWD} (a) and (b) is consistent with that of prior studies \citep{howard1985coronal,howard1986solar,hundhausen1993sizes,st1999comparison,gopalswamy2003coronal,yeh2005kinetic,jang2016comparison}, being somewhat skewed to the right, but displaced towards higher values. \citet{yashiro2004catalog} found that a log-normal distribution well describes the CME speed distribution, and the CME width distribution is also broad and skewed. \citet{gopalswamy2006coronal} studied the speed and width distributions of CMEs observed between 1996 and 2004 and found that the speed distribution is broad and skewed, with a long tail toward higher speeds, and that the width distribution is also broad and skewed, with a peak at around 100 degrees. Our results emphasize that CME parameters must be cautiously handled, particularly when physical quantities such as CME speed and width are considered.

\subsection{Effect of Source Region on the Speed and Width Distributions }
\begin{figure}[h]  
\centering

   \includegraphics[width=\textwidth]{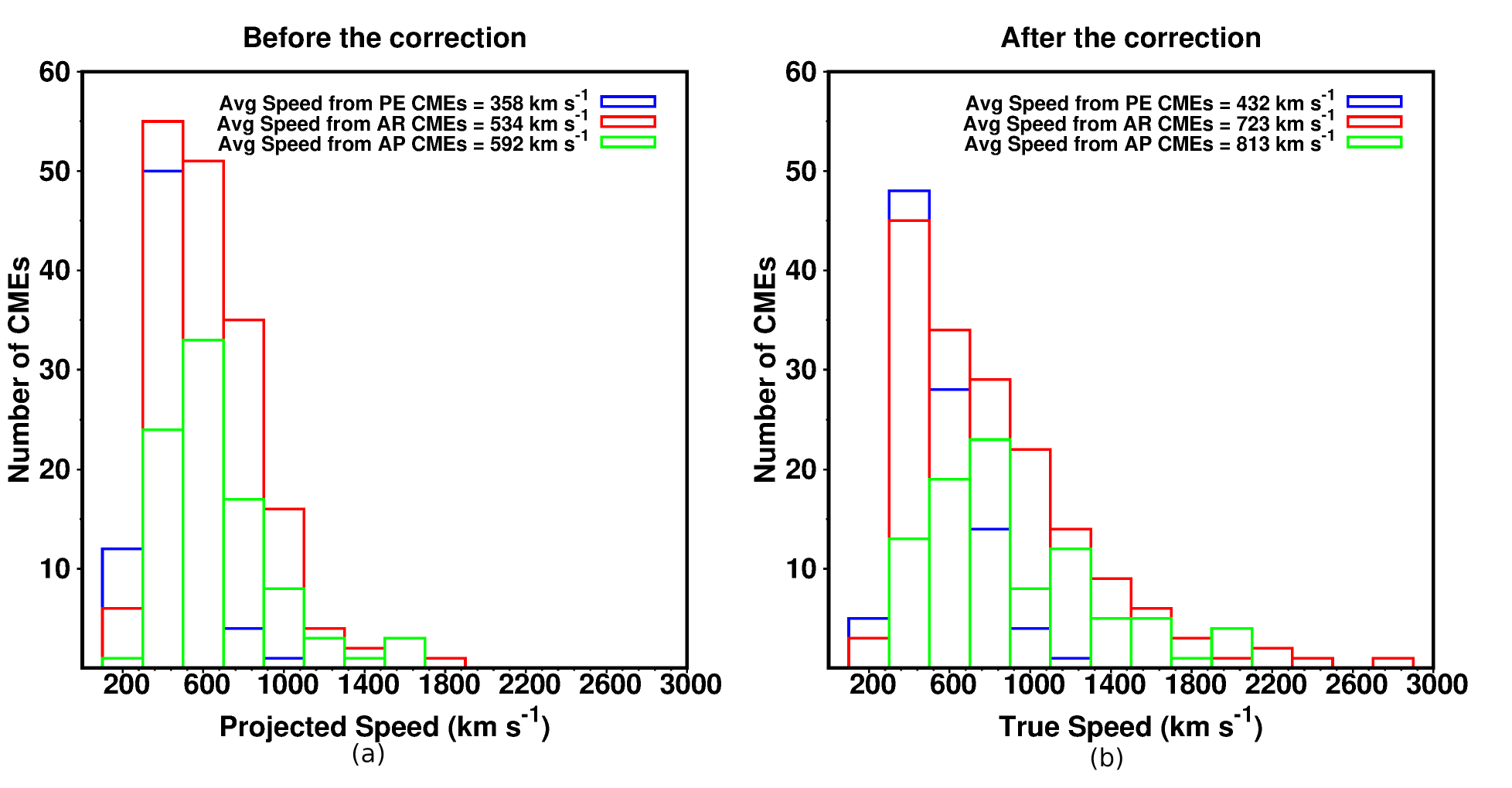}
  
\caption{Histograms of the apparent/projected \textit{(left)} and true \textit{(right)} speed of CMEs based on source region segregation: AR CMEs (shown in red), PE CMEs (shown in blue), and AP CMEs (shown in green) with their corresponding average values.}  
\label{fig:S-W_dist}
\end{figure}

\begin{figure}[h]  
\centering
   \includegraphics[width=\textwidth]{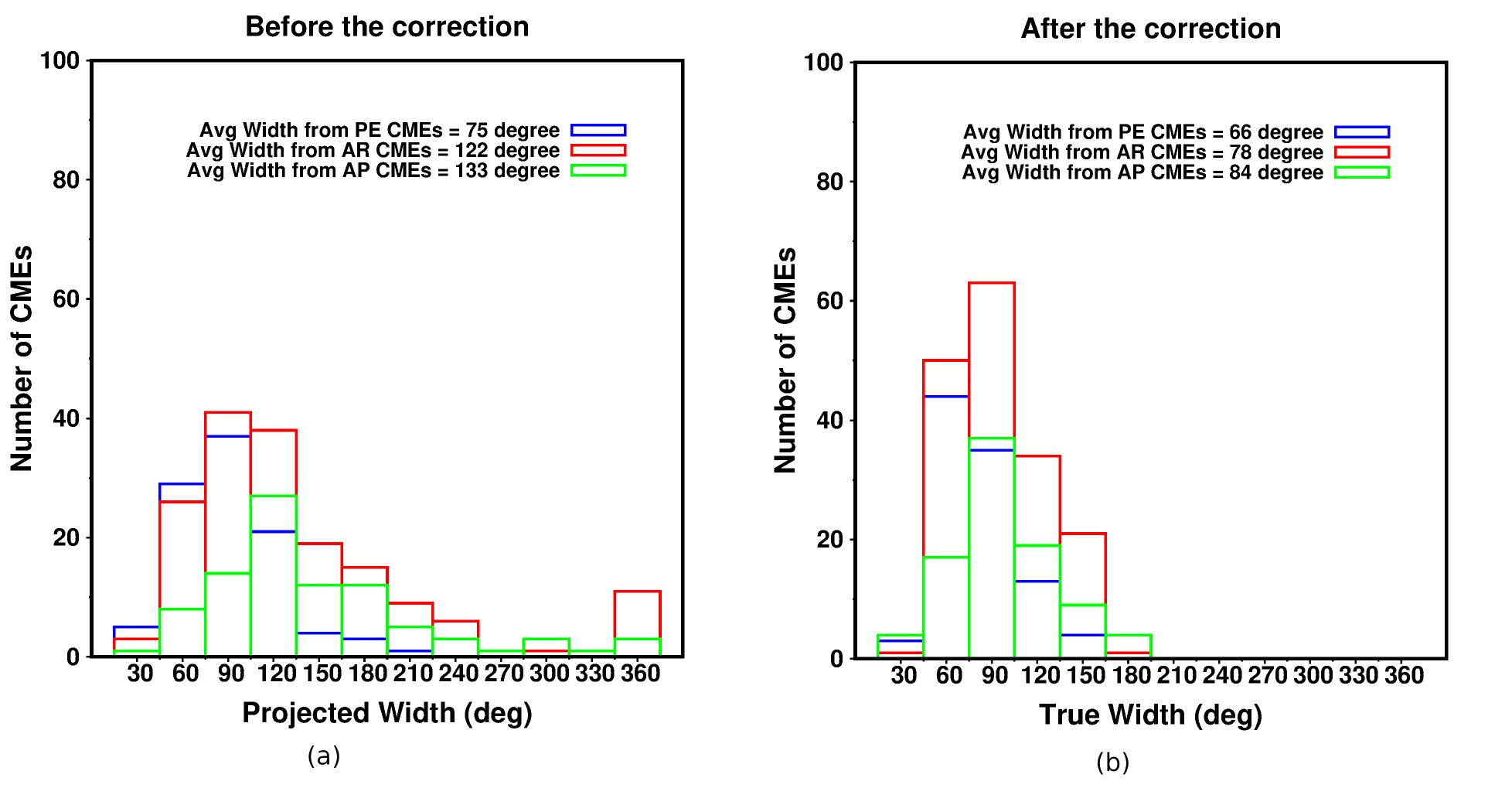}

\caption{Histograms of the width of CMEs before correction \textit{(left)} and after correction \textit{(right)} based on source region segregation: AR CMEs (shown in red), PE CMEs (shown in blue), and AP CMEs (shown in green) with their corresponding average values.}
\label{fig:SR_hist}
\end{figure}

The kinematics of CMEs are heavily influenced by the source region from which they erupt \citep{burlaga1981magnetic,webb2000relationship,subramanian2001source,mishra2005characteristic,zhao2017correlation,pal2018dependence}. Here, we demonstrate how the speed and width distributions (Figures \ref{fig:2D3DSD} and \ref{fig:2D3DWD}) of CMEs change when they are separated into three broad categories: active region eruption \citep{pal2018dependence,majumdar2021insight}, prominence eruption \citep{sheeley1980initial,munro1979association,howard1985coronal,webb1987activity,st1999comparison}, and active prominence eruption \citep{gilbert2000active}. The source region segregated distributions are shown in Figure \ref{fig:S-W_dist} (speed) and Figure \ref{fig:SR_hist} (width). 

Figure \ref{fig:S-W_dist} compares the speed distributions of AR, PE, and AP-CMEs before (left panel) and after (right panel) correcting for projection effects. The CME speeds from AR have the most right-skewed distribution (shown in red), covering a broader range of speeds with 3D speeds ranging from 200-2800  km $s^{-1}$\ (higher than corresponding 2D values) with an average speed of 723  km $s^{-1}$\ as compared to the average 2D speed that is 534  km $s^{-1}$. The CMEs that originate from PE regions (shown in blue), on the other hand, cover a small range of speeds with an average 2D speed of 358  km $s^{-1}$\ and an average 3D speed of 432  km $s^{-1}$. The distribution (shown in green) of CME speeds from the AP region differs significantly from AR and PE distributions, with the highest average 2D speed of 592  km $s^{-1}$ and the highest average 3D speed of 813  km $s^{-1}$. The results in 2D and 3D show that CMEs from PE regions are slow and gradual eruptions as expected. AR distribution has the widest range and very high speeds but with the most probable slow CMEs. APs have a far reduced number of slow CMEs with the highest average speed, suggesting that either the reconnection mechanisms or magnetic environment surrounding the APs are inducive to more impulsive, faster CMEs since footpoints of the eruption are within the region of strong magnetic field, yet the confining magnetic environment may be weaker compared to the active region. It could also be due to such CMEs having a tendency to have a different geometry, perhaps asymmetry, that is poorly modeled by GCS.

Source regions have an effect not only on the speed of CMEs but also on their angular width \citep{zhao2017correlation,pant2021investigating}. Figure \ref{fig:SR_hist} shows the distribution of width influenced by different source region types, before and after correction. Figure \ref{fig:SR_hist} (b) shows that the width distribution of CMEs coming from AR and AP regions are quite similar, with most CMEs of width 90$^{\circ}$ with average widths of 78$^{\circ}$ and 84$^{\circ}$ respectively. CMEs from PE region have an average width of 66$^{\circ}$, peaking at 60$^{\circ}$. This shows that AR and AP-based CMEs are wider than PE CMEs. When compared to their corresponding projected distributions (left panel of Figure \ref{fig:SR_hist}), the average widths are found to be higher in all three cases (effects of projections).

\subsection{CME Speed versus CME Width}
\begin{figure}[h!]
\centering
 
   \includegraphics[width=\textwidth]{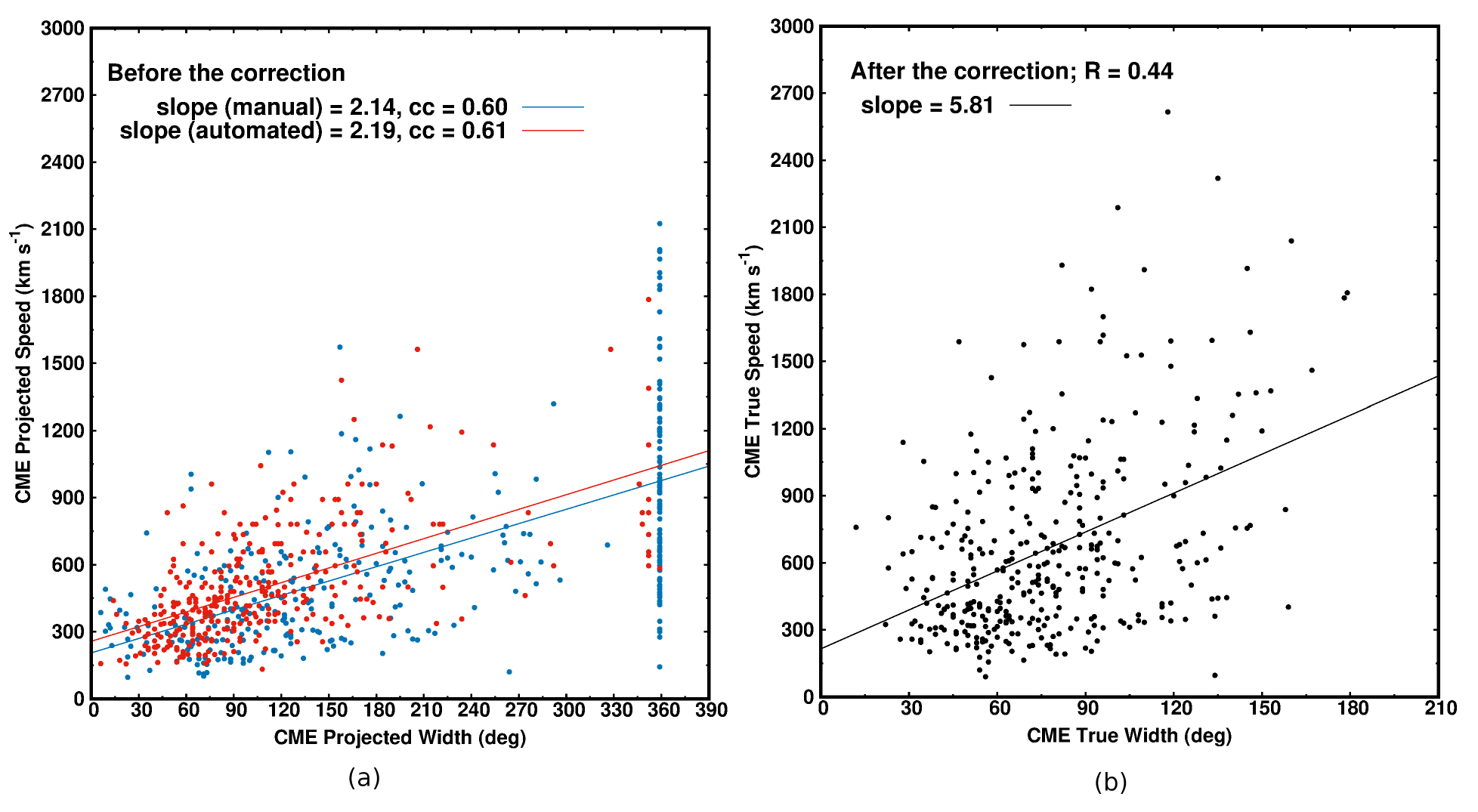}
 
\caption{Scatterplots between CME speed (V) and angular width (W) before correction \textit{(left panel)} and after the correction \textit{(right panel)}. The regression line and the correlation coefficients are indicated on the plot.}
\label{fig:S-W_plot}
\end{figure}

\begin{figure}[h]
    \centering 
   \includegraphics[width=\textwidth]{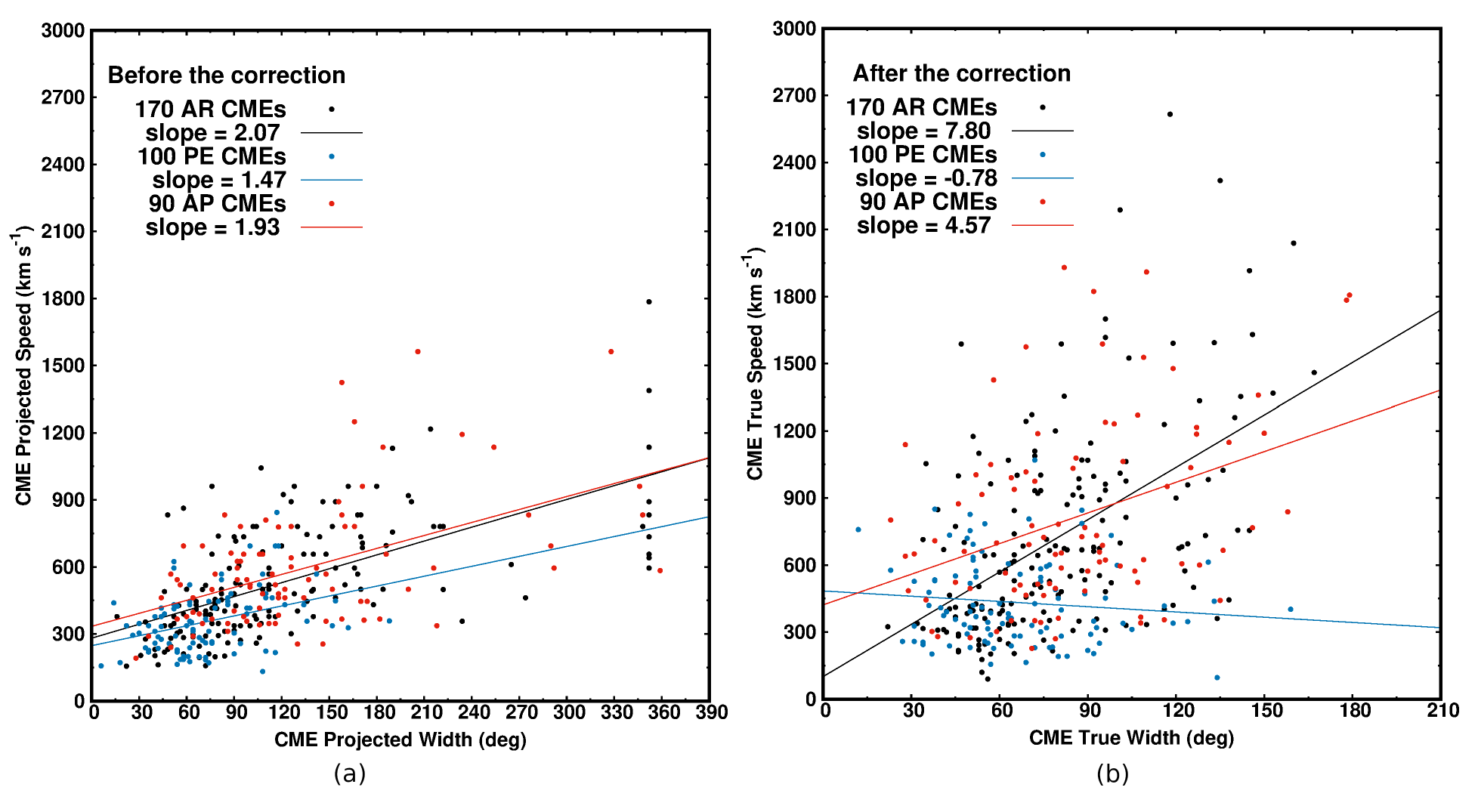}  
\caption{Scatterplots between CME speed (V) and angular width (W) after SR segregation, before and after correction. The regression lines (black for AR-CMEs, red for AP-CMEs, and blue for PE-CMEs) and their correlation coefficients for AR, AP, and PE CMEs are indicated on the plot.}
\label{fig:S-W-SR-plot}
\end{figure}

Wider CMEs tend to be faster than narrower CMEs. This phenomenon has been described in previous studies by \citet{gopalswamy2001characteristics,yashiro2004catalog,burkepile2004role, yeh2005kinetic, vrvsnak2007projection, howard2008kinematic}. These authors demonstrated a weak but positive correlation of $\sim$0.4 between projected speed and width and that the correlation completely disappears after the correction for the projection effect. Note that the parameters used to discover the association in the above studies are projected in the plane-of-sky. With the data sample collected for this paper, the correlation is reexamined. The scatter plots of apparent and true speed plotted as a function of apparent and true width are depicted in figure \ref{fig:S-W_plot}.  We find that the 3D speed-width plot has a weaker correlation (cc = 0.40 and p$<$0.001) and a steeper slope of 5.26 km $s^{-1}\deg^{-1}$ in comparison to their 2D plot, which has cc = 0.60 and a slope of 2.19 km $s^{-1}\deg^{-1}$.
\newpage
\citet{jang2016comparison} conducted a comparison between the 3D and 2D statistical properties of both full and partial halo CMEs. They observed a correlation coefficient of 0.54 between 3D speed and width, which is marginally higher than the correlation found in our study. They noted that the slope in 3D is about five times greater than in 2D when all types of CMEs are considered. However, this increase in slope is approximately 2.5 times for partial halo CMEs alone. This finding aligns with the slope derived from AR CMEs in our study, considering that most halo CMEs originate from AR. \citet{richardson2015properties} and \citet{shen2013full} found a cc = 0.47 and cc = 0.48, which is a little higher than ours (cc = 0.4). A similar correlation can be found in \citet{vrvsnak2007projection} where  the  projected  plane-of-sky  velocity  and angular  width  of  non-halo  CMEs  are  compared.  These differences may be due to the criteria for selecting events. For example, \citet{jang2016comparison} selected CMEs with apparent widths larger than 180, while we selected CMEs ranging in apparent width from 30 to 360. Besides this, we investigated the correlation coefficient and the slope of the speed-width plot before and after correction by looking at their trends from different source regions. Figures \ref{fig:S-W-SR-plot} (a) and (b) show the variation of speed with width for CMEs originating from AR, AP, and PE regions. We found that the slope of the 3D values is almost $\sim$3 times that of their 2D slope in the case of AR-CMEs, and $\sim$2 times in the case of AP-CMEs, implying that wider CMEs originating from ARs tend to be faster as compared to the CMEs of same width but originating from APs. A clear difference can be spotted in the slope for PE-CMEs before and after correction, with a slope of 1.47 (former) and -0.70 (latter), implying almost anticorrelation between the speed and width of CMEs originating from prominences. This relation needs examination with a bigger dataset.

\newpage
\subsection{True Speed versus Projected Speed}

\begin{figure}
    \centering
    \includegraphics[width=0.6\textwidth]{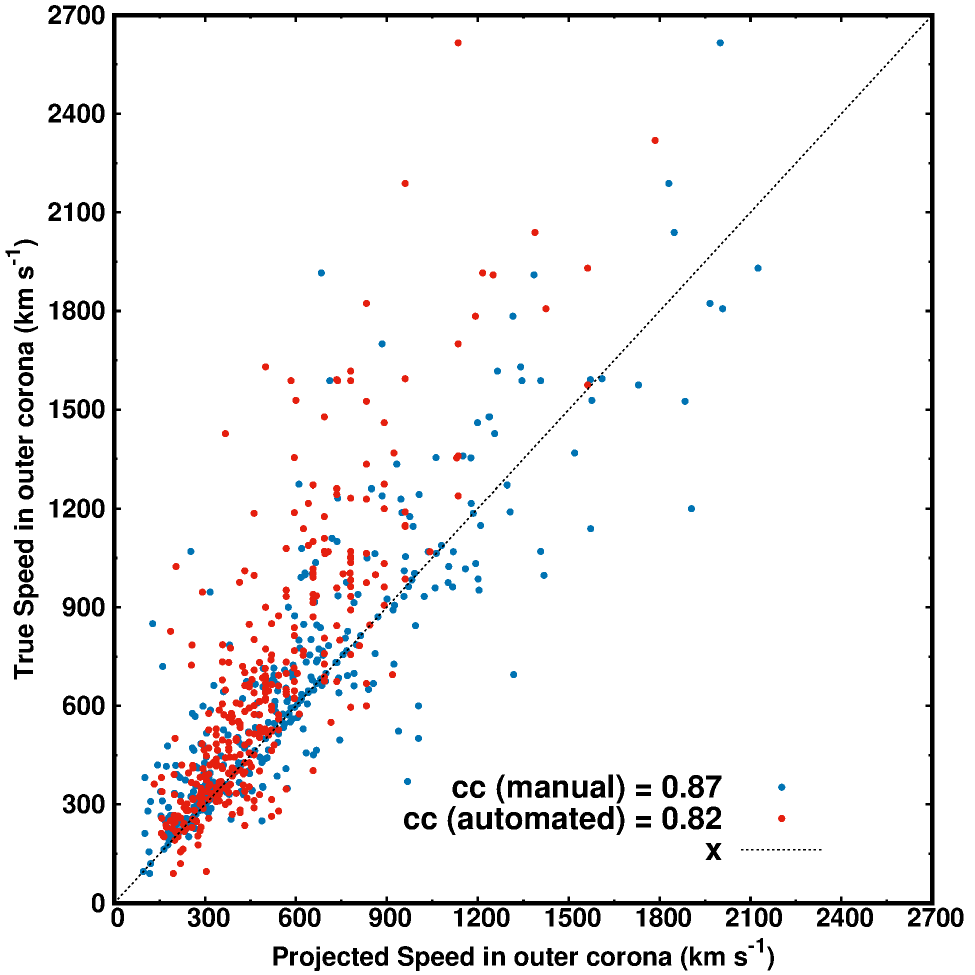}
    \caption{Scatter plot of True speed versus Projected speed. Blue points show 2D speeds in LASCO from manual catalog CDAW, whereas red points show 2D speeds in STEREO-A extracted from automated catalogs CACTus and SEEDs. A strong correlation can be seen in both cases. A dashed line is where the true speed matches the projected speed}
    \label{fig:TPS}
\end{figure}
 Figure \ref{fig:TPS} shows a scatterplot between true speed (on the y-axis), estimated from GCS, and their corresponding projected speed (on the x-axis), taken from CACTus. When the effects of projection are taken into account, the true speeds of all CMEs should ideally be greater than the 2D speeds; however, our findings for a small number of CMEs indicate some inconsistency in this assumption. A linear fitting performed on the two data sets (blue points correspond to projected speeds in STEREO-A, whereas red data points correspond to the projected speed in LASCO FOV) gives a slope of 1.33 and 1.37, which indicates that the true speed varies by a factor of $\sim$1.3 regardless of the coronagraphs that were used to determine the projected speeds. The dashed line is where projected and true speeds are the same.

 \citet{howard2008three} found that the average corrected speed tends to be a factor of 1.7 higher than their projected values. Similar behavior in 3D versus 2D plots can be seen in \citet{majumdar2020connecting} and \citet{jang2016comparison}. A statistical correlation analysis between the corrected and apparent speeds from CACTus and CDAW generated a Pearson’s linear correlation coefficient of cc = 0.82 and 0.87 , showing a strong and positive correlation. Studies (for example \citet{howard2008three,richardson2015properties,jang2016comparison}) found the correlation between the 3D and 2D speed to be cc = 0.71, 0.68, and 0.94.

\section{Summary and Conclusion}
    \label{sec:conclusion}
In prior research, the GCS model has been frequently applied to fit CMEs. The primary catalogs available, KINCAT and CMEDB, extend only up to 2014, halting just as STEREO-A began its transit behind the Sun. As a result, these catalogs mainly capture the ascent of solar cycle 24 and include a restricted number of CMEs evaluated by other scholars. Our dataset is distinct, with no CMEs duplicated and ensuring no redundancy. In our study, we have assembled data on 360 unique CMEs identified by SOHO and STEREO from 2007 to 2021, spanning varied stages of solar cycle 24 with the data gap from October to December in 2014 and January to November in 2015, a period when STEREO-A transitioned behind the Sun. This compilation encompasses 326 GCS-modeled CMEs, assessed by various authors between 2007 and 2014 during the rising phase of solar cycle 24. Our work introduces 34 CMEs that transpired during the declining phase of solar cycle 24 and is a significant addition to the existing datasets. However, we acknowledge the inherent challenges in fully validating the accuracy of these reconstructions, given the current limitations in directly measuring the true properties of CMEs. Combining CME properties from different catalogs could impact the overall results due to the slight variations in the reconstructions of the same CME. These variations often arise because different observers choose different features to fit. A recent paper by \cite{kay2023collection} provides a comprehensive database and gives a detailed discussion on the limitations of variations in the reconstructed parameters across multiple studies. They illustrate a potential variation of up to 19\% in speed and 27\% in angular width in CME reconstructions. Similarly, \cite{majumdar2020connecting} showed a 15\% error in the speed estimation. The uncertainty in speed calculated using the bootstrapping technique for the 34 CMEs fitted during this work falls within these estimated variations. We examined the distributions of speed and width and their interrelationship, which were then associated with their SR type, segregated into three broad categories: ARs, PEs, and APs.  Subsequently, these distributions were compared with the 2D automated parameters cataloged by CACTus (primary catalog) and SEEDS (fill in the missing data) and with the 2D manual parameters cataloged by CDAW to assess the impact of projection. Cactus and SEEDS are the only 2D catalogs that detect CMEs in STEREO images. They often include anomalous and high CME speeds at the position angles adjacent to the flanks, resulting in erroneous CME speed and width measurements. To mitigate these discrepancies, we meticulously reviewed the CACTus catalog and looked at associated movies to obtain the most precise CME speeds. We have also compared the true and projected values acquired via LASCO from the CDAW for 360 events. Both methods yielded similar correlation coefficients (cc) for speed-width distributions (cc=0.60 in the case of CDAW and cc = 0.61 in the case of CACTus). Table \ref{table:AVGCME} summarizes statistical parameters using GCS, CACTus, and CDAW. The main results from this work are:

\begin{enumerate}

    \item Speed distributions of CMEs before and after correction (see Fig. \ref{fig:2D3DSD}) varies significantly, with true speeds ranging from 200-2800  km $s^{-1}$ with an average speed of 665  km $s^{-1}$ as compared to the projected speeds ranging from 200-1800  km $s^{-1}$ with an average speed of 500  km $s^{-1}$, implying that 2D speeds are underestimated by approximately 30\% when compared to 3D ones hence showing a significant amount of projection effects in the 2D estimates.

    \item  Comparison of CME width distributions before and after correction (see Fig. \ref{fig:2D3DWD}) reveals that 2D widths are greatly overestimated due to projection effects compared to their 3D parameters. Following the correction, the number of CME events with wide angular spans (apparent width $>$ 180) drops to zero, while the number of events with narrow angular spans increases significantly. In both histograms, the maximum frequency range peaks around 90$^{\circ}$, with an average true width of 77 and an average projected width of 112$^{\circ}$.
     \newline
    \item True speed and width distributions of CMEs associated with AR, AP, and PE are examined and compared to their apparent distributions (see Fig. \ref{fig:S-W_dist} and \ref{fig:SR_hist}). As expected, CMEs associated with prominences are slow, 
    gradual, and narrower, with an average true speed of 432  km $s^{-1}$\ and average true width of 66$^{\circ}$, whereas CMEs associated with ARs have higher speeds with an average true speed of 723  km $s^{-1}$\ and average width of 78$^{\circ}$, with slower CMEs most likely in both cases (slightly more in PE-CMEs than AR-CMEs).
    \newline
    \item APs have a much smaller number of slow CMEs with the highest average true speed of 813  km $s^{-1}$\ and the highest average width of 84$^{\circ}$, respectively. The values are higher than the average values obtained for CMEs from ARs due to the low number of slow CMEs. This could be due to the reconnection mechanisms or the magnetic environment around the APs causing more sudden, faster CMEs as the footpoints of the eruption are in a region with a stronger magnetic field, yet the confining magnetic field may be weaker compared to an active region.
     \newline
    \item 3D Speed versus width relation (shown in Fig. \ref{fig:S-W_plot}) has a weak but positive correlation cc = 0.40 and p$<$0.001) with wider CMEs having higher speeds with a slope of 5.26 km $s^{-1}\deg^{-1}$. The exact relation in 2D shows a slope of 2.19 km $s^{-1}\deg^{-1}$  with a cc = 0.60. This implies that the correlation between speed and width decreases with the removal of projection effects and that the true behaviour of CMEs is much more complicated when it comes to studying the effect of width on speed. A similar correlation between GCS speed and width is found in a study by \citet{shen2013full}.
     \newline
    \item The relationship between speed and width (shown in Fig. \ref{fig:S-W-SR-plot}) for different SRs has not been extensively studied in earlier studies. A linear fit to CMEs (shown in black and red) follows a trend with a slope of 7.22 (AR) and 3.88 (AP). Almost an anticorrelation is found between speed and width for PE CMEs. Hence, we show that kinematics of CMEs changes significantly when they are associated with different SR types.
    \newline
    

    \item True and projected speeds are compared (see Fig. \ref{fig:TPS}) for 360 CMEs with projected speeds taken from CACTus and CDAW separately. A slope of $\sim$1.3 in both cases implies the true speed is higher than the projected speed regardless of the single viewpoint values. However, a small fraction of CMEs were found to have projected speeds higher than true speeds. A strong and positive correlation of cc = 0.82 and cc = 0.87 can be seen while comparing true speeds with automated and manual projected speeds.
\end{enumerate}

In conclusion, our results show that removing projection effects in CMEs is crucial for accurately understanding their kinematic properties and physical parameters. Several studies have been conducted to investigate the impact of projection effects on CME properties and to develop methods to correct them. In general, eliminating projection effects in CMEs is a crucial step towards gaining an improved understanding of their physical processes and their influence on space weather. However, further research is still needed to improve the existing methods and investigate the impact of CME orientation and width on the speed. Due to the subjectivity of these studies, machine learning techniques such as convolutional neural networks to fit CMEs autonomously in multi-view coronagraph images could be a major step to improve the accuracy of CME speed and width measurements by avoiding the manual bias and reducing the fitting time.

%
%

\section*{Open Research Section}

The data used to make the plots in the paper can be found at \citet{harshu939_2024_cme}

\acknowledgments

HG is supported by an STFC studentship at Aberystwyth University. HM’s contribution is supported by Leverhulme grant RPG-2019-361. We acknowledge SECCHI/STEREO consortium for providing data. The SECCHI data used here were produced by an international consortium of the Naval Research Laboratory (USA), Lockheed Martin Solar and Astrophysics Lab (USA), NASA Goddard Space Flight Center (USA), Rutherford Appleton Laboratory (UK), University of Birmingham (UK), Max-Planck-Institut for Solar System Research (Germany), Centre Spatiale de Li$\grave{e}$ge (Belgium), Institut d'Optique Th$\acute{e}$orique et Appliqu$\acute{e}$e (France), Institut d'Astrophysique Spatiale (France).  We also thank NASA for making SOHO/LASCO data publicly available. SOHO is a project of international cooperation between ESA and NASA. The STEREO/SECCHI/COR2 CME catalog is generated and maintained at the Institute for Astrophysics of the University of Goettingen, supported by the German Space Agency DLR and the European Union in collaboration with the U.S. Naval Research Laboratory, Washington. The authors acknowledge the use of computing resources provided by ARIES and Aberystwyth University.

%
%

\bibliography{sola_bibliography}

\end{document}